# Mass and Heat Diffusion in Ternary Polymer Solutions: A Classical Irreversible Thermodynamics Approach

S. Shams Es-haghi and M. Cakmak[*]

*Department of Polymer Engineering, 250 S. Forge St., The University of Akron, Akron, Ohio, 44325-0301, USA*

**ABSTRACT**

Governing equations for evolution of concentration and temperature in three-component systems were derived in the framework of classical irreversible thermodynamics using Onsager's variational principle and were presented for solvent/solvent/polymer and solvent/polymer/polymer systems. The derivation was developed from the Gibbs equation of equilibrium thermodynamics using the local equilibrium hypothesis, Onsager reciprocal relations and Prigogine's theorem for systems in mechanical equilibrium. It was shown that the details of mass and heat diffusion phenomena in a ternary system are completely expressed by a 3×3 matrix whose entries are mass diffusion coefficients (4 entries), thermal diffusion coefficients (2 entries) and three entries that describe the evolution of heat in the system. The entries of the diffusion matrix are related to the elements of Onsager matrix that are bounded by some constraints to satisfy the positive definiteness of entropy production in the system. All the elements of diffusion matrix were expressed in terms of derivatives of exchange chemical potentials of the components with respect to concentration and temperature. The spinodal curves of ternary polymer solutions were derived from the governing equations and their correctness was checked by the Hessian of free energy density. Moreover, it was proved that setting cross-diffusion coefficients to zero results in a contradiction, and the governing equations without cross-diffusion coefficients do not express the actual phase behavior of the system.

*Corresponding author: cakmak1@uakron.edu

# I. INTRODUCTION

Mass and heat diffusion in polymer solutions is of great importance in the technologies related to painting, coating, inkjet printing, plastic films and production of electronic devices [1]. Diffusion processes also play a key role in self-assembly [2] and important areas of study in soft matter physics [3]. A proper understanding of diffusion phenomena in polymer solutions is required in order to have a control on the processes involving polymer solutions.

Polymer solutions commonly used for industrial purposes are generally multi-component systems, i.e., ternary polymer solutions of one polymer in two solvents or two polymers in one solvent [4]. Most of the experimental and theoretical studies thus far, have been about diffusion in binary systems, and despite the broad range of applications involve ternary polymer solutions [4-14] there is no theoretical work that formulates thermodynamically consistent governing equations for mass and heat diffusion in ternary polymer solutions in a way that all the mass and thermal diffusion coefficients necessary to describe the transport phenomena are derived in terms of thermodynamic variables.

By increasing the number of components in a mixture and emergence of cross-diffusion effects, the transport phenomena become very complex. As a result, a large number of mass diffusion coefficients are required to describe the mass transport [2]. This complexity will increase by adding a spatially varying temperature field to the system, since the effect of thermal diffusion and contributions of mass fluxes to the heat flow must also be considered. Mass and heat diffusion are irreversible phenomena and they contribute to entropy production in the system. Hence, the governing equations for mass and heat evolution in ternary polymer solutions must be derived in the framework of classical irreversible thermodynamics in which the balance of entropy plays a crucial role. An overview of the governing equations formulated for mass and heat diffusion in ternary polymer solutions indicates that these equations were derived without considering the balance of entropy in the system. Vrentas et al. [15] derived governing equations for isothermal mass diffusion in ternary polymer solutions based on a theory due to Bearman [16]. They considered a ternary polymer system containing one polymer and two solvents. In that system, the concentration of solvents was much lower than that of the polymer. They set the cross-diffusion coefficients to zero and estimated the principal mass diffusion coefficients by the self-diffusion coefficients of the solvents. Using the same approach based on the statistical theory of Bearman [16], Shojaie et al. [17] developed a model for non-isothermal mass diffusion in a ternary polymer solution containing a polymer and two solvents. The heat transfer in their model was described by the Fourier's law of heat conduction. Alsoy and Duda [18] studied the drying of ternary polymer solutions using a model in which the mass diffusion coefficients were expressed in terms of self-diffusion coefficients of the solvents. In that model they assumed temperature to be temporally dependent and spatially independent. Dabral et al. [4] modeled the drying process of a ternary polymer solution as an isothermal mass diffusion problem. Due to the lack of reliable experimental data and the absence of a suitable predictive theory, they set the cross-diffusion coefficients to zero. In the above-mentioned formulations, there is no criterion that guarantees the governing equations are thermodynamically consistent. An analysis of these models developed for diffusion in multi-component systems indicates that some of them are not consistent with the Onsager reciprocal relations [19]. Moreover, in these studies, contribution of temperature gradient to mass diffusion in ternary polymer solutions was neglected. Furthermore, the heat transfer was modeled using the Fourier's law of heat conduction. This formulation does not take into account the effect of mass diffusion on the heat transfer and the coupling between mass and heat transfer is only due to the temperature dependency of mass diffusion coefficients.

Recently, we developed a thermodynamically consistent model of mass and heat diffusion in binary polymer solutions [20]. In this paper, we focus on generalization of this model to ternary polymer solutions. Although this study aims to develop a theoretical understanding of diffusion in ternary polymer solutions, the governing equations derived herein can be used for any ternary system. Since, the governing equations have been formulated for the general case of non-isothermal three-component systems, the effect of temperature gradient on the evolution of concentration of the components was shown by derivation of thermal diffusion coefficients in terms of phenomenological coefficients and derivatives of exchange chemical potentials of the components with respect to temperature. Moreover, it was proved that cross-diffusion coefficients play a crucial role in expressing the phase behavior of a ternary system and setting the cross-diffusion coefficients to zero leads to a contradiction.

## II. THEORY

We consider a non-reacting ternary system in which non-convective mass diffusion and heat conduction occur. The system is considered to be in one phase, far from the critical region of phase separation. We also assume the system is in mechanical equilibrium. The hypothesis of local equilibrium allows the fundamental equation of classical thermodynamics to be valid for every volume element in the system, although the whole system is not in equilibrium [21]. Using this hypothesis and Prigogine's theorem for systems in mechanical equilibrium, the rate of change of entropy per unit volume for a three-component system such as solvent/solvent/polymer or solvent/polymer/polymer is given by (see Appendix A)

$$\frac{ds}{dt} = -\nabla \cdot \left( \frac{J_q - \mu_{1,3} J_1 - \mu_{2,3} J_2}{T} \right) - $$
$$\frac{1}{T^2} \nabla T \cdot (J_q - \mu_{1,3} J_1 - \mu_{2,3} J_2) - \frac{1}{T} J_1 \cdot \nabla \mu_{1,3} - \frac{1}{T} J_2 \cdot \nabla \mu_{2,3}, \tag{1}$$

where $s$, $J_q$, $J_1$, $J_2$ and $T$ are entropy per unit volume, heat flux, mass flux of component 1, mass flux of component 2 and absolute temperature, respectively. $\mu_{1,3}$ and $\mu_{2,3}$ are exchange chemical potentials of components 1 and 3, and 2 and 3, respectively.

Equation (1) represents the rate of change of entropy per unit volume of the mixture in terms of divergence of entropy flux and rate of entropy production per unit volume of the system $\sigma$, which can be presented in the bilinear form of thermodynamics forces $\underline{X} = \left( -\frac{1}{T} \nabla \mu_{1,3}, -\frac{1}{T} \nabla \mu_{2,3}, -\frac{1}{T^2} \nabla T \right)$ and the conjugated fluxes $\underline{J} = (J_1, J_2, J_q - \mu_{1,3} J_1 - \mu_{2,3} J_2)$,

$$\sigma = -\frac{1}{T^2} \nabla T \cdot (J_q - \mu_{1,3} J_1 - \mu_{2,3} J_2) - \frac{1}{T} J_1 \cdot \nabla \mu_{1,3} - \frac{1}{T} J_2 \cdot \nabla \mu_{2,3}. \tag{2}$$

As one can see due to a constraint for the fluxes, in a ternary system the mass flux of two components are expressed explicitly in the rate of entropy production. In effect, for a general case of an $n$-component system, $n-1$ equation are required to describe the mass diffusion in the system.

The governing equations can be found using linear flux-force relationships considering the Onsager reciprocal relations and also they can be derived from Onsager's variational principle [22] by maximizing $(\sigma - \psi)_J$, the difference between rate of entropy production in the system and a dissipation function $\psi$ presented in terms of thermodynamic forces in the system

$$\psi(X_1, X_2, X_3) = \frac{1}{2} \sum_{i,k=1}^{3} L_{ik} X_i X_k \geq 0, \tag{3}$$

here $L_{ik}$ are phenomenological coefficients. Using linear flux-force relationships, the fluxes can be written in terms of the forces as shown in Eqs. (4) - (6) which in matrix presentation would yield Eq. (7)

$$J_1 = l_{11}\left(-\frac{1}{T}\nabla\mu_{1,3}\right) + l_{12}\left(-\frac{1}{T}\nabla\mu_{2,3}\right) + l_{13}\left(-\frac{1}{T^2}\nabla T\right), \tag{4}$$

$$J_2 = l_{21}\left(-\frac{1}{T}\nabla\mu_{1,3}\right) + l_{22}\left(-\frac{1}{T}\nabla\mu_{2,3}\right) + l_{23}\left(-\frac{1}{T^2}\nabla T\right), \tag{5}$$

$$J_q - \mu_{1,3}J_1 - \mu_{2,3}J_2 = l_{31}\left(-\frac{1}{T}\nabla\mu_{1,3}\right) + l_{32}\left(-\frac{1}{T}\nabla\mu_{2,3}\right) + l_{33}\left(-\frac{1}{T^2}\nabla T\right), \tag{6}$$

$$\begin{pmatrix} J_1 \\ J_2 \\ J_q - \mu_{1,3}J_1 - \mu_{2,3}J_2 \end{pmatrix} = \begin{pmatrix} l_{11} & l_{12} & l_{13} \\ l_{21} & l_{22} & l_{23} \\ l_{31} & l_{32} & l_{33} \end{pmatrix} \begin{pmatrix} -\frac{1}{T}\nabla\mu_{1,3} \\ -\frac{1}{T}\nabla\mu_{2,3} \\ -\frac{1}{T^2}\nabla T \end{pmatrix}. \tag{7}$$

Entries of matrix $L = (l_{ij})_{3\times 3}$ are Onsager's coefficients and based on Onsager's reciprocity relations, matrix $L$ is a symmetric matrix [23]. In conditions for which linear flux-force relations are valid, rate of entropy production takes the quadratic form

$$\begin{aligned}\sigma = &\, l_{11}\left(-\frac{1}{T}\nabla\mu_{1,3}\right)^2 + l_{22}\left(-\frac{1}{T}\nabla\mu_{2,3}\right)^2 + l_{33}\left(-\frac{1}{T^2}\nabla T\right)^2 + \\ &\, 2l_{12}\left(-\frac{1}{T}\nabla\mu_{1,3}\right)\left(-\frac{1}{T}\nabla\mu_{2,3}\right) + 2l_{13}\left(-\frac{1}{T}\nabla\mu_{1,3}\right)\left(-\frac{1}{T^2}\nabla T\right) + \\ &\, 2l_{23}\left(-\frac{1}{T}\nabla\mu_{2,3}\right)\left(-\frac{1}{T^2}\nabla T\right) > 0.\end{aligned} \tag{8}$$

Matrix $L = (l_{ij})_{3\times 3}$ that satisfies Eq. (8) should be positive definite and to be so, its entries should satisfy the conditions

$$l_{11} > 0, \ l_{11}l_{22} > (l_{12})^2, \ \det \begin{pmatrix} l_{11} & l_{12} & l_{13} \\ l_{12} & l_{22} & l_{23} \\ l_{13} & l_{23} & l_{33} \end{pmatrix} > 0. \tag{9}$$

Since $L$ is a symmetric positive definite matrix, all of its diagonal entries, $l_{ii}$ are positive.

By considering

$$\frac{l_{ii}}{T} = \alpha_i \text{ (where i=1 or 2 or 3)}, \ \frac{l_{12}}{T} = \beta, \ \frac{l_{13}}{T} = \gamma, \ \frac{l_{23}}{T} = \delta, \tag{10}$$

and replacing $\nabla \mu_{1,3}$ and $\nabla \mu_{2,3}$ with the right hand sides of Eqs. (11) and (12), respectively, knowing the fact that chemical potential is a function of concentration and temperature,

$$\nabla \mu_{1,3} = \left( \frac{\partial \mu_{1,3}}{\partial \varphi_1} \right)_{T,\varphi_2} \nabla \varphi_1 + \left( \frac{\partial \mu_{1,3}}{\partial \varphi_2} \right)_{T,\varphi_1} \nabla \varphi_2 + \left( \frac{\partial \mu_{1,3}}{\partial T} \right)_{\varphi_1,\varphi_2} \nabla T, \tag{11}$$

$$\nabla \mu_{2,3} = \left( \frac{\partial \mu_{2,3}}{\partial \varphi_1} \right)_{T,\varphi_2} \nabla \varphi_1 + \left( \frac{\partial \mu_{2,3}}{\partial \varphi_2} \right)_{T,\varphi_1} \nabla \varphi_2 + \left( \frac{\partial \mu_{2,3}}{\partial T} \right)_{\varphi_1,\varphi_2} \nabla T, \tag{12}$$

we can recast Eqs. (4) - (6) in the forms

$$\begin{aligned} J_1 = -&\left[ \alpha_1 \left( \frac{\partial \mu_{1,3}}{\partial \varphi_1} \right)_{T,\varphi_2} + \beta \left( \frac{\partial \mu_{2,3}}{\partial \varphi_1} \right)_{T,\varphi_2} \right] \nabla \varphi_1 - \\ &\left[ \alpha_1 \left( \frac{\partial \mu_{1,3}}{\partial \varphi_2} \right)_{T,\varphi_1} + \beta \left( \frac{\partial \mu_{2,3}}{\partial \varphi_2} \right)_{T,\varphi_1} \right] \nabla \varphi_2 - \\ &\left[ \alpha_1 \left( \frac{\partial \mu_{1,3}}{\partial T} \right)_{\varphi_1,\varphi_2} + \beta \left( \frac{\partial \mu_{2,3}}{\partial T} \right)_{\varphi_1,\varphi_2} + \frac{\gamma}{T} \right] \nabla T, \end{aligned} \tag{13}$$

$$J_2 = -\left[\beta\left(\frac{\partial \mu_{1,3}}{\partial \varphi_1}\right)_{T,\varphi_2} + \alpha_2\left(\frac{\partial \mu_{2,3}}{\partial \varphi_1}\right)_{T,\varphi_2}\right]\nabla\varphi_1 -$$
$$\left[\beta\left(\frac{\partial \mu_{1,3}}{\partial \varphi_2}\right)_{T,\varphi_1} + \alpha_2\left(\frac{\partial \mu_{2,3}}{\partial \varphi_2}\right)_{T,\varphi_1}\right]\nabla\varphi_2 - \qquad(14)$$
$$\left[\beta\left(\frac{\partial \mu_{1,3}}{\partial T}\right)_{\varphi_1,\varphi_2} + \alpha_2\left(\frac{\partial \mu_{2,3}}{\partial T}\right)_{\varphi_1,\varphi_2} + \frac{\delta}{T}\right]\nabla T,$$

$$J_q = \mu_{1,3}J_1 + \mu_{2,3}J_2 - \left[\gamma\left(\frac{\partial \mu_{1,3}}{\partial \varphi_1}\right)_{T,\varphi_2} + \delta\left(\frac{\partial \mu_{2,3}}{\partial \varphi_1}\right)_{T,\varphi_2}\right]\nabla\varphi_1 -$$
$$\left[\gamma\left(\frac{\partial \mu_{1,3}}{\partial \varphi_2}\right)_{T,\varphi_1} + \delta\left(\frac{\partial \mu_{2,3}}{\partial \varphi_2}\right)_{T,\varphi_1}\right]\nabla\varphi_2 - \qquad(15)$$
$$\left[\gamma\left(\frac{\partial \mu_{1,3}}{\partial T}\right)_{\varphi_1,\varphi_2} + \delta\left(\frac{\partial \mu_{2,3}}{\partial T}\right)_{\varphi_1,\varphi_2} + \frac{\alpha_3}{T}\right]\nabla T.$$

In order to preserve the positive definiteness of matrix $L$, the following conditions must be satisfied:

$$\begin{aligned}&\alpha_i > 0 \quad i = 1, 2, 3\\&\alpha_1\alpha_2 - \beta^2 > 0,\\&\alpha_1\alpha_2\alpha_3 + 2\beta\gamma\delta - \alpha_1\delta^2 - \alpha_2\gamma^2 - \alpha_3\beta^2 > 0.\end{aligned} \qquad(16)$$

After deriving the heat and mass fluxes, using the definition of the time derivative of enthalpy per unit volume (see Eq. A.4) and mass concentration for component $i$ (see Eq. A.5), we obtain the governing equations from the divergence of the fluxes

$$\rho_1 \frac{d\varphi_1}{dt} = \nabla \cdot \left\{ \left[ \alpha_1 \left( \frac{\partial \mu_{1,3}}{\partial \varphi_1} \right)_{T,\varphi_2} + \beta \left( \frac{\partial \mu_{2,3}}{\partial \varphi_1} \right)_{T,\varphi_2} \right] \nabla \varphi_1 \right\} +$$
$$\nabla \cdot \left\{ \left[ \alpha_1 \left( \frac{\partial \mu_{1,3}}{\partial \varphi_2} \right)_{T,\varphi_1} + \beta \left( \frac{\partial \mu_{2,3}}{\partial \varphi_2} \right)_{T,\varphi_1} \right] \nabla \varphi_2 \right\} + \quad (17)$$
$$\nabla \cdot \left\{ \left[ \alpha_1 \left( \frac{\partial \mu_{1,3}}{\partial T} \right)_{\varphi_1,\varphi_2} + \beta \left( \frac{\partial \mu_{2,3}}{\partial T} \right)_{\varphi_1,\varphi_2} + \frac{\gamma}{T} \right] \nabla T \right\},$$

$$\rho_2 \frac{d\varphi_2}{dt} = \nabla \cdot \left\{ \left[ \beta \left( \frac{\partial \mu_{1,3}}{\partial \varphi_1} \right)_{T,\varphi_2} + \alpha_2 \left( \frac{\partial \mu_{2,3}}{\partial \varphi_1} \right)_{T,\varphi_2} \right] \nabla \varphi_1 \right\} +$$
$$\nabla \cdot \left\{ \left[ \beta \left( \frac{\partial \mu_{1,3}}{\partial \varphi_2} \right)_{T,\varphi_1} + \alpha_2 \left( \frac{\partial \mu_{2,3}}{\partial \varphi_2} \right)_{T,\varphi_1} \right] \nabla \varphi_2 \right\} + \quad (18)$$
$$\nabla \cdot \left\{ \left[ \beta \left( \frac{\partial \mu_{1,3}}{\partial T} \right)_{\varphi_1,\varphi_2} + \alpha_2 \left( \frac{\partial \mu_{2,3}}{\partial T} \right)_{\varphi_1,\varphi_2} + \frac{\delta}{T} \right] \nabla T \right\},$$

$$\rho c_P \frac{dT}{dt} = -\nabla \cdot \left( \mu_{1,3} J_1 + \mu_{2,3} J_2 \right) + \nabla \cdot \left\{ \left[ \gamma \left( \frac{\partial \mu_{1,3}}{\partial \varphi_1} \right)_{T,\varphi_2} + \delta \left( \frac{\partial \mu_{2,3}}{\partial \varphi_1} \right)_{T,\varphi_2} \right] \nabla \varphi_1 \right\} +$$
$$\nabla \cdot \left\{ \left[ \gamma \left( \frac{\partial \mu_{1,3}}{\partial \varphi_2} \right)_{T,\varphi_1} + \delta \left( \frac{\partial \mu_{2,3}}{\partial \varphi_2} \right)_{T,\varphi_1} \right] \nabla \varphi_2 \right\} + \quad (19)$$
$$\nabla \cdot \left\{ \left[ \gamma \left( \frac{\partial \mu_{1,3}}{\partial T} \right)_{\varphi_1,\varphi_2} + \delta \left( \frac{\partial \mu_{2,3}}{\partial T} \right)_{\varphi_1,\varphi_2} + \frac{\alpha_3}{T} \right] \nabla T \right\},$$

where $\rho$ and $c_P$ are the mass density and isobaric specific heat capacity of the system, respectively.

Using Eqs. (7) and (10), one can express the equation of heat flux in the form (see Appendix B)

$$J_q' = -k\nabla T, \quad (20)$$

where $J_q'$ and $k$ are reduced heat flux and thermal conductivity and are given by

$$J_q' = J_q - \left(\mu_{1,3} + \frac{\alpha_2\gamma - \beta\delta}{\alpha_1\alpha_2 - \beta^2}\right)J_1 - \left(\mu_{2,3} + \frac{\alpha_1\delta - \beta\gamma}{\alpha_1\alpha_2 - \beta^2}\right)J_2, \qquad (21)$$

$$k = \frac{1}{T}\left(\alpha_3 - \frac{\alpha_1\delta^2 - 2\beta\gamma\delta + \alpha_2\gamma^2}{\alpha_1\alpha_2 - \beta^2}\right). \qquad (22)$$

Equation (22) shows thermal conductivity as a function of phenomenological coefficients and the constraint given by Eq. (16) guarantees that the thermal conductivity is a positive quantity.

There is another way of expressing the equation of heat diffusion. This can be done by replacing the mass fluxes in Eq. (19) using Eqs. (13) and (14). Doing so, the equation of heat diffusion will be

$$\begin{aligned}
\rho c_P \frac{dT}{dt} &= \nabla \cdot \left\{\left[(\alpha_1\mu_{1,3} + \beta\mu_{2,3} + \gamma)\left(\frac{\partial\mu_{1,3}}{\partial\varphi_1}\right)_{T,\varphi_2} + (\beta\mu_{1,3} + \alpha_2\mu_{2,3} + \delta)\left(\frac{\partial\mu_{2,3}}{\partial\varphi_1}\right)_{T,\varphi_2}\right]\nabla\varphi_1\right\} + \\
&\quad \nabla \cdot \left\{\left[(\alpha_1\mu_{1,3} + \beta\mu_{2,3} + \gamma)\left(\frac{\partial\mu_{1,3}}{\partial\varphi_2}\right)_{T,\varphi_1} + (\beta\mu_{1,3} + \alpha_2\mu_{2,3} + \delta)\left(\frac{\partial\mu_{2,3}}{\partial\varphi_2}\right)_{T,\varphi_1}\right]\nabla\varphi_2\right\} + \\
&\quad \nabla \cdot \left\{\left[(\alpha_1\mu_{1,3} + \beta\mu_{2,3} + \gamma)\left(\frac{\partial\mu_{1,3}}{\partial T}\right)_{\varphi_1,\varphi_2} + (\beta\mu_{1,3} + \alpha_2\mu_{2,3} + \delta)\left(\frac{\partial\mu_{2,3}}{\partial T}\right)_{\varphi_1,\varphi_2}\right]\nabla T\right\} + \\
&\quad \nabla \cdot \left[\frac{(\gamma\mu_{1,3} + \delta\mu_{2,3} + \alpha_3)}{T}\nabla T\right].
\end{aligned} \qquad (23)$$

Equations (17), (18) and (23) can be written in a compact form as

$$\frac{d}{dt}\begin{pmatrix}\rho_1\varphi_1 \\ \rho_2\varphi_2 \\ \rho c_P T\end{pmatrix} = \nabla \cdot \begin{pmatrix} D_{11} & D_{12} & D_{13} \\ D_{21} & D_{22} & D_{23} \\ D_{31} & D_{32} & D_{33} \end{pmatrix}\begin{pmatrix}\nabla\varphi_1 \\ \nabla\varphi_2 \\ \nabla T\end{pmatrix}. \qquad (24)$$

We call the 3×3 matrix in Eq. (24) diffusion matrix whose entries $D_{ij}$ (see Appendix C) describe the details of mass and heat diffusion in the system. Mass diffusion matrix is a 2×2 matrix with entries $D_{11}$, $D_{12}$, $D_{21}$ and $D_{22}$. $D_{12}$ and $D_{21}$ are cross-diffusion coefficients. $D_{13}$ and $D_{23}$ are thermal diffusion coefficients. The three entries in the third row of the diffusion matrix describe the heat conduction in the system.

## III. RESULTS AND DISCUSSION

In section 2, the mass and heat diffusion coefficients were derived as part of the model in terms of derivatives of exchange chemical potentials of the components with respect to concentration and temperature. Thus, in the model developed herein, there is no need for definition of diffusion coefficients. The approach adopted in this paper for deriving the governing equations in ternary systems can be compared with the model developed by Curtiss and Bird [24] who derived the generalized Maxwell-Stefan equations for the multi-component diffusion. In that model the diffusion coefficients were not derived. Curtiss and Bird [24] considered two different definitions for mass diffusion coefficients; zero-diagonal and symmetric diffusivity definitions and based on these definitions they derived generalized Maxwell-Stephan equations.

The 3D governing equations derived in section 2 can be used for any ternary mixture. The entries of the diffusion matrix will be different depending on the chemical potentials of the components in a particular system. In order to express the governing equations for ternary polymer solutions, one needs to have the chemical potentials of the components. We consider two general ternary polymer solutions; A solution of one polymer in two solvents,

solvent(1)/solvent(2)/polymer(3) and a solution of two polymers in one solvent, solvent(1)/polymer(2)/polymer(3). The numbers in parenthesis will be used as subscripts in equations to represent components in the system. The chemical potentials of the components in a ternary polymer solution can be derived by extension of Flory-Huggins theory to ternary polymer solution [25,26]. Then, the exchange chemical potentials of the components can be derived using Eqs. (A.11) and (A.12).

## A. Diffusion matrix for solvent(1)/solvent(2)/polymer(3) systems

In case of solvent(1)/solvent(2)/polymer(3) systems, the exchange chemical potentials of the components will be (see Appendix D)

$$\frac{M_1(\mu_{1,3} - \mu_{1,3}^0)}{RT} = \ln \varphi_1 - \frac{1}{N_3} \ln \varphi_3 + \left(1 - \frac{1}{N_3}\right) + \left(\chi_{12}\varphi_2 + \chi_{13}\varphi_3\right)\left(\varphi_2 + \varphi_3\right) - \left(\chi_{31}\varphi_1 + \chi_{32}\varphi_2\right)\left(\varphi_1 + \varphi_2\right) - \chi_{23}\varphi_2\varphi_3 + \chi_{12}\varphi_1\varphi_2, \quad (25)$$

$$\frac{M_2(\mu_{2,3} - \mu_{2,3}^0)}{RT} = \ln \varphi_2 - \frac{1}{N_3} \ln \varphi_3 + \left(1 - \frac{1}{N_3}\right) + \left(\chi_{21}\varphi_1 + \chi_{23}\varphi_3\right)\left(\varphi_1 + \varphi_3\right) - \left(\chi_{31}\varphi_1 + \chi_{32}\varphi_2\right)\left(\varphi_1 + \varphi_2\right) - \chi_{13}\varphi_1\varphi_3 + \chi_{12}\varphi_1\varphi_2, \quad (26)$$

where $\mu_{1,3}^0$ and $\mu_{2,3}^0$ are

$$\mu_{1,3}^0 = \frac{1}{M_1}\left(\mu_1^0 - \frac{1}{N_3}\mu_3^0\right), \quad (27)$$

$$\mu_{2,3}^0 = \frac{1}{M_2}\left(\mu_2^0 - \frac{1}{N_3}\mu_3^0\right). \quad (28)$$

Here $\mu_i^0$, $M_1$, $M_2$, $N_3$ and $\chi_{ij}$ are chemical potential of component $i$ in its pure liquid state, molecular weight of component 1, molecular weight of component 2, degree of polymerization of component 3 (polymer) and the Flory-Huggins interaction parameter between components $i$ and $j$.

Using Eqs. (25) – (28) the diffusion matrix for solvent(1)/solvent(2)/polymer(3) systems can be derived (see Appendix E).

**B. Diffusion matrix for solvent(1)/polymer(2)/polymer(3) systems**

In case of solvent(1)/polymer(2)/polymer(3) systems, the exchange chemical potentials of the components will be (see Appendix D)

$$\frac{M_1(\mu_{1,3} - \mu_{1,3}^0)}{RT} = \ln\varphi_1 - \frac{1}{N_3}\ln\varphi_3 + \left(1 - \frac{1}{N_3}\right) + (\chi_{12}\varphi_2 + \chi_{13}\varphi_3)(\varphi_2 + \varphi_3) - \\ (\chi_{31}\varphi_1 + \chi_{32}\varphi_2)(\varphi_1 + \varphi_2) - \chi_{23}\varphi_2\varphi_3 + \chi_{12}\varphi_1\varphi_2, \tag{29}$$

$$\frac{m_2(\mu_{2,3} - \mu_{2,3}^0)}{RT} = \frac{1}{N_2}\ln\varphi_2 - \frac{1}{N_3}\ln\varphi_3 + \left(\frac{1}{N_2} - \frac{1}{N_3}\right) + (\chi_{21}\varphi_1 + \chi_{23}\varphi_3)(\varphi_1 + \varphi_3) - \\ (\chi_{31}\varphi_1 + \chi_{32}\varphi_2)(\varphi_1 + \varphi_2) - \chi_{13}\varphi_1\varphi_3 + \chi_{12}\varphi_1\varphi_2, \tag{30}$$

where $m_2$ and $N_2$ are the molecular weight of the repeating unit of the component 2 (polymer 2) and its degree of polymerization, respectively and $\mu_{1,3}^0$ and $\mu_{2,3}^0$ are given by

$$\mu_{1,3}^0 = \frac{1}{M_1}\left(\mu_1^0 - \frac{1}{N_3}\mu_3^0\right), \tag{31}$$

$$\mu_{2,3}^0 = \frac{1}{M_2}\left(\mu_2^0 - \frac{N_2}{N_3}\mu_3^0\right). \tag{32}$$

Using Eqs. (29) – (32) the diffusion matrix for solvent(1)/polymer(2)/polymer(3) systems can be derived (see Appendix F).

**C. Mass and heat diffusion in an n-component system**

The governing equations derived for ternary mixtures can easily be generalized to describe mass and heat diffusion in an $n$-component system.

The governing equations for mass and heat diffusion in an $n$-component system are given by

$$\frac{d}{dt}\begin{pmatrix} \rho_1\varphi_1 \\ \vdots \\ \rho_{(n-1)}\varphi_{(n-1)} \\ \rho c_P T \end{pmatrix} = \nabla \cdot \left( \begin{array}{ccc|c} \begin{bmatrix} D_{11} & \cdots & D_{1(n-1)} \\ \vdots & \ddots & \vdots \\ D_{(n-1)1} & \cdots & D_{(n-1)(n-1)} \end{bmatrix} & \begin{pmatrix} D_{1n} \\ \vdots \\ D_{(n-1)n} \end{pmatrix} \\ \begin{pmatrix} D_{n1} & \cdots & D_{n(n-1)} \end{pmatrix} & D_{nn} \end{array} \right) \begin{pmatrix} \nabla\varphi_1 \\ \vdots \\ \nabla\varphi_{(n-1)} \\ \nabla T \end{pmatrix}. \quad (33)$$

The $(n-1)\times(n-1)$ submatrix shown inside a box in Eq. (33) is the mass diffusion matrix and the components of the vector in the last column of diffusion matrix are thermal diffusion coefficients. The Onsager matrix associated with the diffusion matrix is an $n\times n$ symmetric positive definite matrix whose entries must satisfy

$$l_{11} > 0, \; l_{11}l_{22} > (l_{12})^2, \; \det\begin{pmatrix} l_{11} & l_{12} & l_{13} \\ l_{12} & l_{22} & l_{23} \\ l_{13} & l_{23} & l_{33} \end{pmatrix} > 0, \; \ldots, \det\begin{pmatrix} l_{11} & \cdots & l_{1n} \\ \vdots & \ddots & \vdots \\ l_{1n} & \cdots & l_{nn} \end{pmatrix} > 0. \quad (34)$$

This concept of generalizing governing equations to an $n$-component system may become important when a solution of several polymers is prepared in one or several solvents. Moreover, knowing the fact that polymers have a distribution of molecular weight, a solution of one polymer in one solvent is a multi-component system of polymer chains with different molecular weights in that solvent [28].

**D. Derivation of spinodal curve from the governing equations.**

Important information about phase behavior of a system can be extracted from the governing equations for mass diffusion in the system. The spinodal hypersurface for an $n$-components system is given by [29]

$$\det\begin{pmatrix} D_{11} & \cdots & D_{1(n-1)} \\ \vdots & \ddots & \vdots \\ D_{(n-1)1} & \cdots & D_{(n-1)(n-1)} \end{pmatrix} = 0. \qquad (35)$$

In effect, the spinodal curve for a system is the hypersurface on which the mass diffusion matrix becomes singular. Therefore, for a ternary mixture the spinodal curve is given by

$$\det\begin{pmatrix} \alpha_1\left(\dfrac{\partial \mu_{1,3}}{\partial \varphi_1}\right)_{T,\varphi_2} + \beta\left(\dfrac{\partial \mu_{2,3}}{\partial \varphi_1}\right)_{T,\varphi_2} & \alpha_1\left(\dfrac{\partial \mu_{1,3}}{\partial \varphi_2}\right)_{T,\varphi_1} + \beta\left(\dfrac{\partial \mu_{2,3}}{\partial \varphi_2}\right)_{T,\varphi_1} \\ \beta\left(\dfrac{\partial \mu_{1,3}}{\partial \varphi_1}\right)_{T,\varphi_2} + \alpha_2\left(\dfrac{\partial \mu_{2,3}}{\partial \varphi_1}\right)_{T,\varphi_2} & \beta\left(\dfrac{\partial \mu_{1,3}}{\partial \varphi_2}\right)_{T,\varphi_1} + \alpha_2\left(\dfrac{\partial \mu_{2,3}}{\partial \varphi_2}\right)_{T,\varphi_1} \end{pmatrix} = 0. \qquad (36)$$

Due to the constraint $\alpha_1\alpha_2 - \beta^2 > 0$, the determinant of the mass diffusion matrix will be

$$\left(\dfrac{\partial \mu_{1,3}}{\partial \varphi_1}\right)_{T,\varphi_2}\left(\dfrac{\partial \mu_{2,3}}{\partial \varphi_2}\right)_{T,\varphi_1} - \left(\dfrac{\partial \mu_{1,3}}{\partial \varphi_2}\right)_{T,\varphi_1}\left(\dfrac{\partial \mu_{2,3}}{\partial \varphi_1}\right)_{T,\varphi_2} = 0. \qquad (37)$$

Equation (37) gives the spinodal curve for a general ternary system. Using the equations of exchange chemical potentials of the components, the spinodal curve for solvent(1)/solvent(2)/polymer(3) and solvent(1)/polymer(2)/polymer(3) systems will be

$$\left(\dfrac{1}{\varphi_1} + \dfrac{1}{N_3(1-\varphi_3)} - 2\chi_{13}\right)\left(\dfrac{1}{\varphi_2} + \dfrac{1}{N_3(1-\varphi_3)} - 2\chi_{23}\right) - \left(\dfrac{1}{N_3(1-\varphi_3)} + \chi_{12} - \chi_{13} - \chi_{23}\right)^2 = 0, \qquad (38)$$

$$\left(\dfrac{1}{\varphi_1} + \dfrac{1}{N_3(1-\varphi_3)} - 2\chi_{13}\right)\left(\dfrac{1}{N_2\varphi_2} + \dfrac{1}{N_3(1-\varphi_3)} - 2\chi_{23}\right) - \left(\dfrac{1}{N_3(1-\varphi_3)} + \chi_{12} - \chi_{13} - \chi_{23}\right)^2 = 0, \qquad (39)$$

respectively. The results derived herein can be checked with a well-established alternative way in that the spinodal hypersurface of an $n$-component system can be derived from the determinant of the Hessian matrix of free energy density of the system [3,30]

$$\det \begin{pmatrix} \dfrac{\partial^2 f}{\partial \varphi_1^2} & \cdots & \dfrac{\partial^2 f}{\partial \varphi_1 \partial \varphi_{(n-1)}} \\ \vdots & \ddots & \vdots \\ \dfrac{\partial^2 f}{\partial \varphi_{(n-1)} \partial \varphi_1} & \cdots & \dfrac{\partial^2 f}{\partial \varphi_{(n-1)}^2} \end{pmatrix} = 0. \tag{40}$$

Therefore, for a ternary mixture the spinodal curve will be

$$\left( \frac{\partial^2 f}{\partial \varphi_1^2} \right) \left( \frac{\partial^2 f}{\partial \varphi_2^2} \right) - \left( \frac{\partial^2 f}{\partial \varphi_1 \partial \varphi_2} \right) \left( \frac{\partial^2 f}{\partial \varphi_2 \partial \varphi_1} \right) = 0. \tag{41}$$

Free energy density functions of solvent(1)/solvent(2)/polymer(3) and solvent(1)/polymer(2)/polymer(3) systems are given by [31]

$$f = \varphi_1 \ln \varphi_1 + \varphi_2 \ln \varphi_2 + \frac{\varphi_3 \ln \varphi_3}{N_3} + \chi_{12} \varphi_1 \varphi_2 + \chi_{13} \varphi_1 \varphi_3 + \chi_{23} \varphi_2 \varphi_3, \tag{42}$$

$$f = \varphi_1 \ln \varphi_1 + \frac{\varphi_2 \ln \varphi_2}{N_2} + \frac{\varphi_3 \ln \varphi_3}{N_3} + \chi_{12} \varphi_1 \varphi_2 + \chi_{13} \varphi_1 \varphi_3 + \chi_{23} \varphi_2 \varphi_3, \tag{43}$$

respectively. The free energy density functions are presented in dimensionless form and according to Eq. (41) this does not affect the result of the problem. It is easy to show that the spinodal curves derived from Eq. (41) using Eqs. (42) and (43) are identical with the spinodal curves derived from the governing equations.

### E. The consequence of setting cross-diffusion coefficients to zero

According to the literature (see for example Ref. 3), due to the lack of data or in order to simplify the problem the cross-diffusion coefficients are set to zero. However, it has been observed that dropping cross-diffusion coefficients resulted in significant errors in predictions of drying of ternary polymer solutions [32]. In this section we discuss the consequence of ignoring cross-diffusion coefficients in a ternary system.

By dropping the cross-diffusion coefficient, .i.e. $D_{12} = D_{21} = 0$, the mass diffusion matrix for a ternary system will be presented either in the form

$$\begin{pmatrix} \left(\beta - \dfrac{\alpha_1 \alpha_2}{\beta}\right)\left(\dfrac{\partial \mu_{2,3}}{\partial \varphi_1}\right)_{T,\varphi_2} & 0 \\ 0 & \left(\alpha_2 - \dfrac{\beta^2}{\alpha_1}\right)\left(\dfrac{\partial \mu_{2,3}}{\partial \varphi_2}\right)_{T,\varphi_1} \end{pmatrix}, \quad (44)$$

or

$$\begin{pmatrix} \left(\alpha_1 - \dfrac{\beta^2}{\alpha_2}\right)\left(\dfrac{\partial \mu_{1,3}}{\partial \varphi_1}\right)_{T,\varphi_2} & 0 \\ 0 & \left(\beta - \dfrac{\alpha_2 \alpha_1}{\beta}\right)\left(\dfrac{\partial \mu_{1,3}}{\partial \varphi_2}\right)_{T,\varphi_1} \end{pmatrix}. \quad (45)$$

Due to the constraint $\alpha_1 \alpha_2 - \beta^2 > 0$, the spinodal curve of the system will be

$$\left(\dfrac{\partial \mu_{2,3}}{\partial \varphi_1}\right)_{T,\varphi_2} \left(\dfrac{\partial \mu_{2,3}}{\partial \varphi_2}\right)_{T,\varphi_1} = 0, \quad (46)$$

or

$$\left(\dfrac{\partial \mu_{1,3}}{\partial \varphi_1}\right)_{T,\varphi_2} \left(\dfrac{\partial \mu_{1,3}}{\partial \varphi_2}\right)_{T,\varphi_1} = 0, \quad (47)$$

respectively. It is clear that the spinodal curve demonstrated by the governing equations after reducing the cross-diffusion coefficients to zero is not the actual spinodal curve of the system. Moreover, Eqs. (46) and (47) lead to a contradiction in that the exchange chemical potentials of the components $\mu_{1,3}$ and $\mu_{2,3}$ must be a constant or be only a function of $\varphi_1$ or $\varphi_2$ so that Eqs.

(46) and (47) are valid. Therefore, one can see that setting the off-diagonal entries of mass diffusion matrix to zero can result in a serious problem in that not only the actual phase behavior of the system will be altered, but also a contradiction arises. Therefore, one is not eligible to ignore the cross-diffusion coefficients in the system. It is worth mentioning that the proof provided herein is for a general case and can be considered for any three-component system.

## IV. SUMMARY AND CONCLUSIONS

3D governing equations for mass and heat diffusion in three-component systems were derived in the framework of classical irreversible thermodynamics. The formulation of the governing equations was developed from the Gibbs equation using Onsager's variation principle. It was shown that the details of mass and heat diffusion in a three-component mixture are described by a 3×3 diffusion matrix whose entries are mass diffusion coefficients, thermal diffusion coefficients and three entries that describe the evolution of temperature field in the system. Mass diffusion coefficients are the entries of mass diffusion matrix, a 2×2 submatrix which remains after elimination of the third row and third column of the diffusion matrix. The entries of the diffusion matrix were expressed in terms of phenomenological coefficients and derivatives of exchange chemical potentials of the components with respect to concentration and temperature. The model was formulated for the general case of non-isothermal mass diffusion in three-component systems and can be used for any three-component system, when the chemical potentials of the components are known as functions of concentration and temperature. The entries of the diffusion matrix were derived for particular cases of ternary polymer solutions, solvent/solvent/polymer and solvent/polymer/polymer systems by exploiting the Flory-Huggins theory to derive the chemical potentials of the components. It was shown that the model can be easily generalized to describe mass and heat diffusion in an $n$-component system. In this general case diffusion matrix has $n^2$ entries where $(n-1)^2$ entries are mass diffusion coefficients, $(n-1)$ entries are thermal diffusion coefficients and $n$ entries describe the evolution of temperature field in the system.

The spinodal curves of ternary polymer solutions were derived from the entries of the mass diffusion matrix and their correctness was checked using the determinant of the Hessian matrix of free energy density functions of the systems. Moreover, for general case of ternary systems, it was shown that cross-diffusion coefficients play a crucial role in expressing the phase behavior of the system. In effect, it was proved that setting the cross-diffusion coefficients to zero leads to a contradiction.

**ACKNOWLEDGMENT**

We are grateful for funding entitled "Polyimide research and Commercialization Grant" from The State of Ohio Third frontier program.

**Appendix A. Rate of change of entropy per unit volume of a ternary system**

Equation (A.1) states the fundamental equation per unit volume of a mixture;

$$de = Tds + \sum_{i=1}^{k} \mu_i dn_i, \qquad (A.1)$$

where $T$, $s$, $e$, $\mu_i$ and $n_i$ are absolute temperature, entropy per unit volume, enthalpy per unit volume, chemical potential of component $i$ and moles of component $i$ per unit volume, respectively.

One can rewrite Eq. (A.1), using the mass concentration of the components and take a derivative with respect to time to find

$$T\frac{ds}{dt} = \frac{de}{dt} - \sum_{i=1}^{k} \frac{\mu_i}{M_i} \frac{dc_i}{dt}, \qquad (A.2)$$

where $c_i$ and $M_i$ are the mass concentration and molecular weight of component $i$, respectively.

Time derivatives in Eq. (A.2) are substantial time derivatives given by

$$\frac{d}{dt} = \frac{\partial}{\partial t} + v \cdot \nabla. \qquad (A.3)$$

Here, $v$ is the mean velocity of the components. The time derivatives of $e$ and $c_i$ in Eq. (A.2) are related to the divergence of heat and non-convective mass fluxes,

$$\frac{de}{dt} = -\nabla \cdot J_q, \qquad (A.4)$$

$$\frac{dc_i}{dt} = -\nabla \cdot J_i. \qquad (A.5)$$

According to Prigogine's theorem, for systems in mechanical equilibrium an arbitrary frame of reference can be chosen [21]. The non-convective mass flux of the component $i$ in this frame of reference which moves with the mean velocity $v$ is given by

$$J_i = \rho_i \varphi_i (v_i - v), \tag{A.6}$$

where $\rho_i$, $\varphi_i$ and $v_i$ are mass density, volume fraction and the velocity of component $i$, respectively and the mean velocity $v$ is expressed as

$$v = \sum_{i=1}^{k} \varphi_i v_i. \tag{A.7}$$

It follows from Eqs. (A.6) and (A.7) that the fluxes are not independent and their dependency is given by

$$\sum_{i=1}^{k} \frac{J_i}{\rho_i} = 0. \tag{A.8}$$

Rewriting Eq. (A.2) for a three-component system and replacing $\frac{de}{dt}$ and $\frac{dc_i}{dt}$ with the divergences of the associated fluxes given by Eqs. (A.4) and (A.5), we obtain

$$T \frac{ds}{dt} = -\nabla \cdot J_q + \frac{\mu_1}{M_1} \nabla \cdot J_1 + \frac{\mu_2}{M_2} \nabla \cdot J_2 + \frac{\mu_3}{M_3} \nabla \cdot J_3 \tag{A.9}$$

The subscripts 1, 2 and 3 are attributed to component 1, component 2 and component 3 in the system, respectively. Equation (A.9) can be reduced after applying the constraint for fluxes given by Eq. (A.8) to the form

$$T \frac{ds}{dt} = -\nabla \cdot J_q + \mu_{1,3} \nabla \cdot J_1 + \mu_{2,3} \nabla \cdot J_2, \tag{A.10}$$

where $\mu_{1,3}$ and $\mu_{2,3}$ are the exchange chemical potentials of the components and are given by

$$\mu_{1,3} = \frac{\mu_1}{M_1} - \frac{\rho_3}{\rho_1}\frac{\mu_3}{M_3}, \tag{A.11}$$

$$\mu_{2,3} = \frac{\mu_2}{M_2} - \frac{\rho_3}{\rho_2}\frac{\mu_3}{M_3}, \tag{A.12}$$

respectively. We can rewrite Eq. (A.10) by replacing the right hand side with an equivalent form and dividing both sides by $T$,

$$\frac{ds}{dt} = -\frac{1}{T}\nabla \cdot \left(J_q - \mu_{1,3}J_1 - \mu_{2,3}J_2\right) - \frac{1}{T}J_1 \cdot \nabla\mu_{1,3} - \frac{1}{T}J_2 \cdot \nabla\mu_{2,3}. \tag{A.13}$$

The first term on the right hand side of Eq. (A.13) can be manipulated to an equivalent form so that

$$\begin{aligned}\frac{ds}{dt} = -\nabla \cdot \left(\frac{J_q - \mu_{1,3}J_1 - \mu_{2,3}J_2}{T}\right) - \\ \frac{1}{T^2}\nabla T \cdot \left(J_q - \mu_{1,3}J_1 - \mu_{2,3}J_2\right) - \frac{1}{T}J_1 \cdot \nabla\mu_{1,3} - \frac{1}{T}J_2 \cdot \nabla\mu_{2,3}.\end{aligned} \tag{A.14}$$

## Appendix B. Derivation of reduced heat flux

Using Eq. (10), and knowing that $l_{12} = l_{21} = \beta T$, $l_{13} = l_{31} = \gamma T$, and $l_{23} = l_{32} = \delta T$ the linear flux-force relationships given by Eq. (7) can be written as

$$J_1 = -\alpha_1 \nabla \mu_{1,3} - \beta \nabla \mu_{2,3} - \frac{\gamma}{T} \nabla T, \tag{B.1}$$

$$J_2 = -\beta \nabla \mu_{1,3} - \alpha_2 \nabla \mu_{2,3} - \frac{\delta}{T} \nabla T, \tag{B.2}$$

$$J_q = \mu_{1,3} J_1 + \mu_{2,3} J_2 - \gamma \nabla \mu_{1,3} - \delta \nabla \mu_{2,3} - \frac{\alpha_3}{T} \nabla T, \tag{B.3}$$

Using Eqs. (B.1) and (B.2), $\nabla \mu_{1,3}$ and $\nabla \mu_{2,3}$ are given by

$$\nabla \mu_{1,3} = \left( \frac{-\alpha_2}{\alpha_1 \alpha_2 - \beta^2} \right) J_1 + \left( \frac{\beta}{\alpha_1 \alpha_2 - \beta^2} \right) J_2 + \left( \frac{\beta \delta - \alpha_2 \gamma}{\alpha_1 \alpha_2 - \beta^2} \right) \frac{1}{T} \nabla T, \tag{B.4}$$

$$\nabla \mu_{2,3} = \left( \frac{\beta}{\alpha_1 \alpha_2 - \beta^2} \right) J_1 - \left( \frac{\alpha_1}{\alpha_1 \alpha_2 - \beta^2} \right) J_2 - \left( \frac{\alpha_1 \delta - \beta \gamma}{\alpha_1 \alpha_2 - \beta^2} \right) \frac{1}{T} \nabla T. \tag{B.5}$$

By substituting for $\nabla \mu_{1,3}$ and $\nabla \mu_{2,3}$ in Eq. (B.3), the heat flux will be

$$J_q = \left( \mu_{1,3} + \frac{\alpha_2 \gamma - \beta \delta}{\alpha_1 \alpha_2 - \beta^2} \right) J_1 + \left( \mu_{2,3} + \frac{\alpha_1 \delta - \beta \gamma}{\alpha_1 \alpha_2 - \beta^2} \right) J_2 - \frac{1}{T} \left( \alpha_3 - \frac{\alpha_1 \delta^2 - 2\beta \gamma \delta + \alpha_2 \gamma^2}{\alpha_1 \alpha_2 - \beta^2} \right) \nabla T. \tag{B.6}$$

Equation (B.6) can be expressed in terms of the reduced heat flux that is given by

$$J_q' = J_q - \left( \mu_{1,3} + \frac{\alpha_2 \gamma - \beta \delta}{\alpha_1 \alpha_2 - \beta^2} \right) J_1 - \left( \mu_{2,3} + \frac{\alpha_1 \delta - \beta \gamma}{\alpha_1 \alpha_2 - \beta^2} \right) J_2, \tag{B.7}$$

Therefore, we obtain

$$J_q' = -\frac{1}{T} \left( \alpha_3 - \frac{\alpha_1 \delta^2 - 2\beta \gamma \delta + \alpha_2 \gamma^2}{\alpha_1 \alpha_2 - \beta^2} \right) \nabla T. \tag{B.8}$$

## Appendix C. Entries of diffusion matrix

$$D_{11} = \alpha_1 \left(\frac{\partial \mu_{1,3}}{\partial \varphi_1}\right)_{T,\varphi_2} + \beta \left(\frac{\partial \mu_{2,3}}{\partial \varphi_1}\right)_{T,\varphi_2},$$

$$D_{12} = \alpha_1 \left(\frac{\partial \mu_{1,3}}{\partial \varphi_2}\right)_{T,\varphi_1} + \beta \left(\frac{\partial \mu_{2,3}}{\partial \varphi_2}\right)_{T,\varphi_1},$$

$$D_{13} = \alpha_1 \left(\frac{\partial \mu_{1,3}}{\partial T}\right)_{\varphi_1,\varphi_2} + \beta \left(\frac{\partial \mu_{2,3}}{\partial T}\right)_{\varphi_1,\varphi_2} + \frac{\gamma}{T},$$

$$D_{21} = \beta \left(\frac{\partial \mu_{1,3}}{\partial \varphi_1}\right)_{T,\varphi_2} + \alpha_2 \left(\frac{\partial \mu_{2,3}}{\partial \varphi_1}\right)_{T,\varphi_2},$$

$$D_{22} = \beta \left(\frac{\partial \mu_{1,3}}{\partial \varphi_2}\right)_{T,\varphi_1} + \alpha_2 \left(\frac{\partial \mu_{2,3}}{\partial \varphi_2}\right)_{T,\varphi_1},$$

$$D_{23} = \beta \left(\frac{\partial \mu_{1,3}}{\partial T}\right)_{\varphi_1,\varphi_2} + \alpha_2 \left(\frac{\partial \mu_{2,3}}{\partial T}\right)_{\varphi_1,\varphi_2} + \frac{\delta}{T},$$

$$D_{31} = (\alpha_1 \mu_{1,3} + \beta \mu_{2,3} + \gamma) \left(\frac{\partial \mu_{1,3}}{\partial \varphi_1}\right)_{T,\varphi_2} + (\beta \mu_{1,3} + \alpha_2 \mu_{2,3} + \delta) \left(\frac{\partial \mu_{2,3}}{\partial \varphi_1}\right)_{T,\varphi_2},$$

$$D_{32} = (\alpha_1 \mu_{1,3} + \beta \mu_{2,3} + \gamma) \left(\frac{\partial \mu_{1,3}}{\partial \varphi_2}\right)_{T,\varphi_1} + (\beta \mu_{1,3} + \alpha_2 \mu_{2,3} + \delta) \left(\frac{\partial \mu_{2,3}}{\partial \varphi_2}\right)_{T,\varphi_1},$$

$$D_{33} = (\alpha_1 \mu_{1,3} + \beta \mu_{2,3} + \gamma) \left(\frac{\partial \mu_{1,3}}{\partial T}\right)_{\varphi_1,\varphi_2} + (\beta \mu_{1,3} + \alpha_2 \mu_{2,3} + \delta) \left(\frac{\partial \mu_{2,3}}{\partial T}\right)_{\varphi_1,\varphi_2} + \frac{(\gamma \mu_{1,3} + \delta \mu_{2,3} + \alpha_3)}{T}.$$

**Appendix D. Exchange chemical potentials of components in a ternary polymer solution**

Exchange chemical potentials $\mu_{1,3}$ and $\mu_{2,3}$ can be derived using Eqs. (A.11) and (A.12) which can be written in the forms

$$\mu_{1,3} = \frac{1}{M_1}\left(\mu_1 - \frac{\overline{V_1}}{\overline{V_3}}\mu_3\right), \quad \mu_{2,3} = \frac{1}{M_2}\left(\mu_2 - \frac{\overline{V_2}}{\overline{V_3}}\mu_3\right), \tag{D.1}$$

where $\overline{V_i}/\overline{V_j}$ is the ratio of molar volumes of the components and will be considered equal to the ratio of their degrees of polymerization.

Chemical potentials of the components in a ternary polymer solution such as solvent(1)/solvent(2)/polymer(3) with respect to their pure states are given by

$$\frac{\mu_1 - \mu_1^0}{RT} = \ln\varphi_1 + \left(1 - \frac{1}{N_3}\right)\varphi_3 + (\chi_{12}\varphi_2 + \chi_{13}\varphi_3)(\varphi_2 + \varphi_3) - \chi_{23}\varphi_2\varphi_3, \tag{D.2}$$

$$\frac{\mu_2 - \mu_2^0}{RT} = \ln\varphi_2 + \left(1 - \frac{1}{N_3}\right)\varphi_3 + (\chi_{21}\varphi_1 + \chi_{23}\varphi_3)(\varphi_1 + \varphi_3) - \chi_{13}\varphi_1\varphi_3, \tag{D.3}$$

$$\frac{\mu_3 - \mu_3^0}{RT} = \ln\varphi_3 + (1 - N_3)\varphi_1 + (1 - N_3)\varphi_2 + \\ (\chi_{31}N_3\varphi_1 + \chi_{32}N_3\varphi_2)(\varphi_1 + \varphi_2) - \chi_{12}N_3\varphi_1\varphi_2, \tag{D.4}$$

where $\mu_i^0$, $R$, $N_3$ and $\chi_{ij}$ are chemical potentials of component $i$ in its pure liquid state, gas constant, degree of polymerization of component 3 (polymer) and the Flory-Huggins interaction parameter of components $i$ and $j$. Therefore, the exchange chemical potentials for the solvent(1)/solvent(2)/polymer(3) systems will be

$$\frac{M_1(\mu_{1,3} - \mu_{1,3}^0)}{RT} = \ln\varphi_1 - \frac{1}{N_3}\ln\varphi_3 + \left(1 - \frac{1}{N_3}\right) + (\chi_{12}\varphi_2 + \chi_{13}\varphi_3)(\varphi_2 + \varphi_3) - \\ (\chi_{31}\varphi_1 + \chi_{32}\varphi_2)(\varphi_1 + \varphi_2) - \chi_{23}\varphi_2\varphi_3 + \chi_{12}\varphi_1\varphi_2, \tag{D.5}$$

$$\frac{M_2(\mu_{2,3} - \mu_{2,3}^0)}{RT} = \ln \varphi_2 - \frac{1}{N_3}\ln \varphi_3 + \left(1 - \frac{1}{N_3}\right) + (\chi_{21}\varphi_1 + \chi_{23}\varphi_3)(\varphi_1 + \varphi_3) -$$
$$(\chi_{31}\varphi_1 + \chi_{32}\varphi_2)(\varphi_1 + \varphi_2) - \chi_{13}\varphi_1\varphi_3 + \chi_{12}\varphi_1\varphi_2, \quad \text{(D.6)}$$

where $\mu_{1,3}^0$ and $\mu_{2,3}^0$ are

$$\mu_{1,3}^0 = \frac{1}{M_1}\left(\mu_1^0 - \frac{1}{N_3}\mu_3^0\right), \quad \text{(D.7)}$$

$$\mu_{2,3}^0 = \frac{1}{M_2}\left(\mu_2^0 - \frac{1}{N_3}\mu_3^0\right). \quad \text{(D.8)}$$

Chemical potentials of the components in a ternary polymer solution such as solvent(1)/polymer(2)/polymer(3) with respect to their pure states are given by

$$\frac{\mu_1 - \mu_1^0}{RT} = \ln \varphi_1 + \left(1 - \frac{1}{N_2}\right)\varphi_2 + \left(1 - \frac{1}{N_3}\right)\varphi_3 +$$
$$(\chi_{12}\varphi_2 + \chi_{13}\varphi_3)(\varphi_2 + \varphi_3) - \chi_{23}\varphi_2\varphi_3, \quad \text{(D.9)}$$

$$\frac{\mu_2 - \mu_2^0}{RT} = \ln \varphi_2 + (1 - N_2)\varphi_1 + \left(1 - \frac{N_2}{N_3}\right)\varphi_3 +$$
$$(\chi_{21}N_2\varphi_1 + \chi_{23}N_2\varphi_3)(\varphi_1 + \varphi_3) - \chi_{13}N_2\varphi_1\varphi_3, \quad \text{(D.10)}$$

$$\frac{\mu_3 - \mu_3^0}{RT} = \ln \varphi_3 + (1 - N_3)\varphi_1 + \left(1 - \frac{N_3}{N_2}\right)\varphi_2 +$$
$$(\chi_{31}N_3\varphi_1 + \chi_{32}N_3\varphi_2)(\varphi_1 + \varphi_2) - \chi_{12}N_3\varphi_1\varphi_2. \quad \text{(D.11)}$$

Therefore, the exchange chemical potentials for the solvent(1)/polymer(2)/polymer(3) systems will be

$$\frac{M_1(\mu_{1,3} - \mu_{1,3}^0)}{RT} = \ln \varphi_1 - \frac{1}{N_3}\ln \varphi_3 + \left(1 - \frac{1}{N_3}\right) + (\chi_{12}\varphi_2 + \chi_{13}\varphi_3)(\varphi_2 + \varphi_3) -$$
$$(\chi_{31}\varphi_1 + \chi_{32}\varphi_2)(\varphi_1 + \varphi_2) - \chi_{23}\varphi_2\varphi_3 + \chi_{12}\varphi_1\varphi_2, \quad \text{(D.12)}$$

$$\frac{m_2(\mu_{2,3}-\mu_{2,3}^0)}{RT} = \frac{1}{N_2}\ln\varphi_2 - \frac{1}{N_3}\ln\varphi_3 + \left(\frac{1}{N_2}-\frac{1}{N_3}\right) + (\chi_{21}\varphi_1 + \chi_{23}\varphi_3)(\varphi_1+\varphi_3) -$$
$$(\chi_{31}\varphi_1 + \chi_{32}\varphi_2)(\varphi_1+\varphi_2) - \chi_{13}\varphi_1\varphi_3 + \chi_{12}\varphi_1\varphi_2,$$
(D.13)

where $m_2$ and $N_2$ are the molecular weight of the repeating unit of the component 2 (polymer 2) and its degree of polymerization, respectively and $\mu_{1,3}^0$ and $\mu_{2,3}^0$ are given by

$$\mu_{1,3}^0 = \frac{1}{M_1}\left(\mu_1^0 - \frac{1}{N_3}\mu_3^0\right),$$
(D.14)

$$\mu_{2,3}^0 = \frac{1}{M_2}\left(\mu_2^0 - \frac{N_2}{N_3}\mu_3^0\right).$$
(D.15)

**Appendix E. Entries of diffusion matrix for solvent(1)/solvent(2)/polymer(3) systems**

Since the chemical potentials of pure components are only functions of temperature [27], the derivative of $\mu_i^0$ with respect to the concentration vanishes. Moreover, the interaction parameters were treated as constants. In the case where a temperature dependency is considered for interaction parameters the terms containing derivatives with respect to the temperature must be modified.

$$D_{11} = \frac{\alpha_1 RT}{M_1}\left(\frac{1}{\varphi_1}+\frac{1}{N_3(1-\varphi_3)}-2\chi_{13}\right)+\frac{\beta RT}{M_2}\left(\frac{1}{N_3(1-\varphi_3)}+\chi_{21}-\chi_{23}-\chi_{13}\right),$$

$$D_{12} = \frac{\alpha_1 RT}{M_1}\left(\frac{1}{N_3(1-\varphi_3)}+\chi_{12}-\chi_{13}-\chi_{23}\right)+\frac{\beta RT}{M_2}\left(\frac{1}{\varphi_2}+\frac{1}{N_3(1-\varphi_3)}-2\chi_{23}\right),$$

$$D_{13} = \alpha_1\left[\frac{d\mu_{1,3}^0}{dT}+\frac{(\mu_{1,3}-\mu_{1,3}^0)}{T}\right]+\beta\left[\frac{d\mu_{2,3}^0}{dT}+\frac{(\mu_{2,3}-\mu_{2,3}^0)}{T}\right]+\frac{\gamma}{T},$$

$$D_{21} = \frac{\beta RT}{M_1}\left(\frac{1}{\varphi_1}+\frac{1}{N_3(1-\varphi_3)}-2\chi_{13}\right)+\frac{\alpha_2 RT}{M_2}\left(\frac{1}{N_3(1-\varphi_3)}+\chi_{21}-\chi_{23}-\chi_{13}\right),$$

$$D_{22} = \frac{\beta RT}{M_1}\left(\frac{1}{N_3(1-\varphi_3)}+\chi_{12}-\chi_{13}-\chi_{23}\right)+\frac{\alpha_2 RT}{M_2}\left(\frac{1}{\varphi_2}+\frac{1}{N_3(1-\varphi_3)}-2\chi_{23}\right),$$

$$D_{23} = \beta\left[\frac{d\mu_{1,3}^0}{dT}+\frac{(\mu_{1,3}-\mu_{1,3}^0)}{T}\right]+\alpha_2\left[\frac{d\mu_{2,3}^0}{dT}+\frac{(\mu_{2,3}-\mu_{2,3}^0)}{T}\right]+\frac{\delta}{T},$$

$$D_{31} = \frac{RT(\alpha_1\mu_{1,3}+\beta\mu_{2,3}+\gamma)}{M_1}\left(\frac{1}{\varphi_1}+\frac{1}{N_3(1-\varphi_3)}-2\chi_{13}\right)+$$
$$\frac{RT(\beta\mu_{1,3}+\alpha_2\mu_{2,3}+\delta)}{M_2}\left(\frac{1}{N_3(1-\varphi_3)}+\chi_{21}-\chi_{23}-\chi_{13}\right),$$

$$D_{32} = \frac{RT(\alpha_1 \mu_{1,3} + \beta \mu_{2,3} + \gamma)}{M_1} \left( \frac{1}{N_3(1-\varphi_3)} + \chi_{12} - \chi_{13} - \chi_{23} \right) +$$

$$\frac{RT(\beta \mu_{1,3} + \alpha_2 \mu_{2,3} + \delta)}{M_2} \left( \frac{1}{\varphi_2} + \frac{1}{N_3(1-\varphi_3)} - 2\chi_{23} \right),$$

$$D_{33} = (\alpha_1 \mu_{1,3} + \beta \mu_{2,3} + \gamma) \left[ \frac{d\mu_{1,3}^0}{dT} + \frac{(\mu_{1,3} - \mu_{1,3}^0)}{T} \right] +$$

$$(\beta \mu_{1,3} + \alpha_2 \mu_{2,3} + \delta) \left[ \frac{d\mu_{2,3}^0}{dT} + \frac{(\mu_{2,3} - \mu_{2,3}^0)}{T} \right] + \frac{(\gamma \mu_{1,3} + \delta \mu_{2,3} + \alpha_3)}{T}.$$

**Appendix F. Entries of diffusion matrix for solvent(1)/polymer(2)/polymer(3) systems**

Since the chemical potentials of pure components are only functions of temperature [27], the derivative of $\mu_i^0$ with respect to the concentration vanishes. Moreover, the interaction parameters were treated as constants. In the case where a temperature dependency is considered for interaction parameters the terms containing derivatives with respect to the temperature must be modified.

$$D_{11} = \frac{\alpha_1 RT}{M_1}\left(\frac{1}{\varphi_1} + \frac{1}{N_3(1-\varphi_3)} - 2\chi_{13}\right) + \frac{\beta RT}{m_2}\left(\frac{1}{N_3(1-\varphi_3)} + \chi_{21} - \chi_{23} - \chi_{13}\right),$$

$$D_{12} = \frac{\alpha_1 RT}{M_1}\left(\frac{1}{N_3(1-\varphi_3)} + \chi_{12} - \chi_{13} - \chi_{23}\right) + \frac{\beta RT}{m_2}\left(\frac{1}{N_2\varphi_2} + \frac{1}{N_3(1-\varphi_3)} - 2\chi_{23}\right),$$

$$D_{13} = \alpha_1\left[\frac{d\mu_{1,3}^0}{dT} + \frac{(\mu_{1,3} - \mu_{1,3}^0)}{T}\right] + \beta\left[\frac{d\mu_{2,3}^0}{dT} + \frac{(\mu_{2,3} - \mu_{2,3}^0)}{T}\right] + \frac{\gamma}{T},$$

$$D_{21} = \frac{\beta RT}{M_1}\left(\frac{1}{\varphi_1} + \frac{1}{N_3(1-\varphi_3)} - 2\chi_{13}\right) + \frac{\alpha_2 RT}{m_2}\left(\frac{1}{N_3(1-\varphi_3)} + \chi_{21} - \chi_{23} - \chi_{13}\right),$$

$$D_{22} = \frac{\beta RT}{M_1}\left(\frac{1}{N_3(1-\varphi_3)} + \chi_{12} - \chi_{13} - \chi_{23}\right) + \frac{\alpha_2 RT}{m_2}\left(\frac{1}{N_2\varphi_2} + \frac{1}{N_3(1-\varphi_3)} - 2\chi_{23}\right),$$

$$D_{23} = \beta\left[\frac{d\mu_{1,3}^0}{dT} + \frac{(\mu_{1,3} - \mu_{1,3}^0)}{T}\right] + \alpha_2\left[\frac{d\mu_{2,3}^0}{dT} + \frac{(\mu_{2,3} - \mu_{2,3}^0)}{T}\right] + \frac{\delta}{T},$$

$$D_{31} = \frac{RT(\alpha_1\mu_{1,3} + \beta\mu_{2,3} + \gamma)}{M_1}\left(\frac{1}{\varphi_1} + \frac{1}{N_3(1-\varphi_3)} - 2\chi_{13}\right) +$$
$$\frac{RT(\beta\mu_{1,3} + \alpha_2\mu_{2,3} + \delta)}{m_2}\left(\frac{1}{N_3(1-\varphi_3)} + \chi_{21} - \chi_{23} - \chi_{13}\right),$$

$$D_{32} = \frac{RT(\alpha_1\mu_{1,3} + \beta\mu_{2,3} + \gamma)}{M_1}\left(\frac{1}{N_3(1-\varphi_3)} + \chi_{12} - \chi_{13} - \chi_{23}\right) +$$

$$\frac{RT(\beta\mu_{1,3} + \alpha_2\mu_{2,3} + \delta)}{m_2}\left(\frac{1}{N_2\varphi_2} + \frac{1}{N_3(1-\varphi_3)} - 2\chi_{23}\right),$$

$$D_{33} = (\alpha_1\mu_{1,3} + \beta\mu_{2,3} + \gamma)\left[\frac{d\mu_{1,3}^0}{dT} + \frac{(\mu_{1,3} - \mu_{1,3}^0)}{T}\right] +$$

$$(\beta\mu_{1,3} + \alpha_2\mu_{2,3} + \delta)\left[\frac{d\mu_{2,3}^0}{dT} + \frac{(\mu_{2,3} - \mu_{2,3}^0)}{T}\right] + \frac{(\gamma\mu_{1,3} + \delta\mu_{2,3} + \alpha_3)}{T}.$$

# NOTATION

| | |
|---|---|
| $e$ | enthalpy per unit volume |
| $T$ | absolute temperature |
| $s$ | entropy per unit volume |
| $\mu_i$ | chemical potential of component $i$ |
| $\mu_i^0$ | chemical potential of component $i$ in its pure state |
| $n_i$ | moles of component $i$ per unit volume |
| $t$ | time |
| $M_i$ | molecular weight of component $i$ |
| $c_i$ | mass concentration of component $i$ |
| $J_q$ | heat flux |
| $J_i$ | non-convective mass flux of component $i$ |
| $\rho_i$ | mass density of component $i$ |
| $\varphi_i$ | volume fraction of component $i$ |
| $v_i$ | velocity of component $i$ |
| $v$ | mean velocity |
| $\mu_{i,j}$ | exchange chemical potential of components $i$ and $j$ |
| $\sigma$ | rate of entropy production per unit volume |
| $\psi$ | dissipation function |
| $L$ | Onsager matrix |
| $\underline{X}$ | vector of thermodynamic forces |
| $\underline{J}$ | vector of thermodynamic fluxes |
| $\alpha_i, \beta, \gamma, \delta$ | phenomenological coefficients |
| $\rho$ | mass density of the system |
| $c_P$ | isobaric specific heat capacity of the system |
| $J_q'$ | reduced heat flux |
| $k$ | thermal conductivity |
| $D_{ij}$ | entries of diffusion matrix |
| $\chi_{ij}$ | Flory-Huggins interaction parameter of components $i$ and $j$ |
| $R$ | gas constant |
| $f$ | free energy density function |
| $N_i$ | degree of polymerization of component $i$ |
| $m_2$ | molecular weight of repeating unit of component 2 |
| $\bar{V}_i$ | molar volume of component $i$ |